\documentclass{iopart}
\newcommand\text[1]{\ensuremath{\mathrm{#1}}}
\newcommand\dyad[1] {\ensuremath{\mathrm{#1}}}
\usepackage{bm}
\usepackage{ upgreek }
\usepackage{graphicx}
\renewcommand{\vec}[1]{\boldsymbol{\mathbf{#1}}}

\newcommand\ppar{\ensuremath{p_\parallel}}
\newcommand\pperp{\ensuremath{p_\perp}}

\newcommand\kpar{\ensuremath{k_\parallel}}
\newcommand\kperp{\ensuremath{k_\perp}}
\newcommand\nunit{\ensuremath{\hat{\vec{n}}}}
\newcommand\bunit{\ensuremath{\hat{\vec{b}}}}
\newcommand\Unit[1]{\ensuremath{\hat{\vec{#1}}}}

\newcommand\Div[1]{\ensuremath{\nabla\cdot#1}}
\newcommand\Curl[1]{\ensuremath{\nabla\times#1}}

\newcommand\Equation[1]{Eq.~(\ref{#1})}

\newcommand\Eqns[2]{Eqs.~(#1)~and~(#2)}
\newcommand\EqnsMany[2]{Eqs.~(#1)-(#2)}

\newcommand\Norm[1]{\ensuremath{\|#1\|}}

\newcommand\Deby[1]{\ensuremath{\frac{d}{d#1}}}

\newcommand\Brackets[1]{\ensuremath{\left(#1\right)}}
\newcommand\SBrackets[1]{\ensuremath{\left[#1\right]}}

\newcommand\Leftnabla{\nabla_\leftarrow}
\newcommand\Partby[1]{\ensuremath{\frac{\partial}{\partial#1}}}
\newcommand\Fullby[1]{\ensuremath{\frac{d}{d#1}}}
\newcommand\Bsq{\ensuremath{\Norm{\vec{B}_0}^2}}

\newcommand\Alfven{Alfv\'{e}n}

\begin{document}
%\begin{frontmatter}
\title{Magnetohydrodynamic normal mode analysis of plasma with equilibrium pressure anisotropy}
\date{\today}
\begin{abstract}
In this work, we generalise linear magnetohydrodynamic (MHD) stability theory to include equilibrium pressure anisotropy in the fluid part of the analysis. A novel `single-adiabatic' (SA) fluid closure is presented which is complementary to the usual `double-adiabatic' (CGL) model and has the advantage of naturally reproducing exactly the MHD spectrum in the isotropic limit. As with MHD and CGL, the SA model neglects the anisotropic perturbed pressure and thus loses non-local fast-particle stabilisation present in the kinetic approach. Another interesting aspect of this new approach is that the stabilising terms appear naturally as separate viscous corrections leaving the isotropic SA closure unchanged.  After verifying the self-consistency of the SA model, we re-derive the projected linear MHD set of equations required for stability analysis of tokamaks in the MISHKA code. The cylindrical wave equation is derived analytically as done previously in the spectral theory of MHD and clear predictions are made for the modification to fast-magnetosonic and slow ion sound speeds due to equilibrium anisotropy.
\end{abstract}
\author{M. Fitzgerald}
\address{CCFE
Culham Science Centre,
Abingdon,
Oxon, OX14 3DB, UK}
\ead{Michael.Fitzgerald@ccfe.ac.uk}
\author{M. J. Hole, Z. S. Qu}
\address{Research School of Physics and Engineering, the Australian National University, Canberra ACT 0200, Australia}
%\end{frontmatter}
\section{Introduction}
A hydrodynamic plasma model provides an intuitive and simple framework for plasma dynamics where the details of the velocity distribution of constituent charged particles is not important to the phenomena being studied. In laboratory fusion plasma, the magnetohydrodynamic (MHD) equations successfully explain a range of plasma instabilities. The equations specify necessary conditions for well behaved toroidal magnetic confinement experiments, and provide insights into the low-frequency electromagnetic wave observations, such as the shear \Alfven~wave. 

The stability analysis of tokamak plasma is often broken down into contributions from thermal and non-thermal populations, with fluid models convenient for the thermal components and kinetic models deployed for the rest. The subject of equilibrium pressure anisotropy and flow on tokamaks has received some recent attention due to the upgrading of heating power and the lower reliance on Joule/`Ohmic' heating in recent and future tokamak experiments, particularly in tight aspect ratio (for example: MAST-U and NSTX-U). Codes for MHD equilibrium \cite{Lao1985, HUYSMANS1991,Hirshman1991} are being generalised \cite{Fitzgerald2013,Qu2014,Cooper2009} to include some of these non-resonant non-thermal effects in the fluid part of the analysis. The natural question is how to self-consistently include the stability aspects of pressure anisotropy in a fluid framework. Of course, kinetic extensions to the MHD model for non-Maxwellian parts of collisionless plasmas are widely implemented (see \cite{Lauber2013} for a recent review), however where resonant wave-particle interactions are not important, the full description of the velocity distribution inherent in a kinetic approach should ideally be simplified to a reduced model containing localised `fluid-only effects' to aid in the understanding. One method to make this simplification is with a physically defensible, or at least comprehensible, statement about the per species fluid thermodynamics leading to a simple closure, such as the ideal gas adiabatic pressure law.  Alternatively, one may arbitrarily truncate the hierarchy of fluid equations at a sufficiently high level as to hopefully capture all the non-resonant physics of local heat flow, if perhaps at the cost of clarity of physical interpretation \cite{Ramos2005}.

For collisionless plasma, Chew, Goldberger and Low (CGL) \cite{Chew1956} deduced that the lowest order form of the plasma pressure may be expressed in terms of pressures parallel and perpendicular to the field. They further defined a fluid closure, the double-adiabatic model, which assumed that both pressures did work in their respective directions but neglecting any heat flow between the degrees of freedom, or from adjacent fluid elements. Subsequent analytical work on plasma stability \cite{Bernstein1958} included both MHD and CGL versions of their expressions, with CGL giving the more optimistic stability threshold. This was followed soon after by comparisons of the stability predictions of these fluid models to the kinetic theory \cite{Kruskal1958}\cite{Rosenbluth1959} which, for isotropic equilibrium, revealed that the reason for the difference between fluid models lay in the energy contribution from differing perturbed parallel and perpendicular pressures, with CGL exaggerating stability and MHD underestimating it when compared to the kinetic results. Much subsequent work has been done to analyse the stabilising effects of anisotropic equilibrium \cite{Spies1974,Connor1976,Choe1977}, particularly with regard to the ballooning mode \cite{Rosenbluth1983,Cooper1981,Wang1990,Cheng1994,Bishop1985}, and the internal kink \cite{Graves2003,Graves2005}. Investigation of the stable spectrum in the presence of equilibrium anisotropy is much more sparse \cite{Volkov1966}.

The theoretical shortcomings of the fluid models are well publicised for magnetically confined fusion (see for example \cite{Freidberg1982,Kulsrud1983}) primarily due to the length scales parallel to the magnetic field being too short when compared with the collisional mean free path. Nevertheless, the self-consistent and theoretically rich MHD model serves as a well-understood foundation for understanding fusion plasma, on which one may later build if and when the experimental evidence demands.

This work concerns using a fluid approach for linear stability analysis of anisotropic tokamaks. We describe a novel `single-adiabatic' (SA) fluid closure for the stability analysis of anisotropic magnetised plasma equilibrium, a closure which has the unique property of producing the same results as the MHD model for isotropic equilibria. We re-derive the set of equations currently used in the MISHKA \cite{ISI:A1997YB84300003} code showing the changes required to implement anisotropy, using either single or double adiabatic approaches, but focusing mostly on the single-adiabatic approach. Finally, we apply this new model analytically to cylindrical geometry and make quantitative and falsifiable predictions about the affect of anisotropic equilibrium on the continuous MHD spectrum and sound speed for finite $\beta$.
\section{Single-adiabatic model}
For convenience and clarity, we proceed in a natural MHD unit system where $\mu_0\to 1$. All electromagnetic fields, fluxes and vector potentials may be restored to S.I. units with a transformation $... \to .../\sqrt{\mu_0}$, all electric currents with $j\to\sqrt{\mu_0}j$ and the electrical conductivity with $\sigma \to \mu_0 \sigma$. The model is founded on the quasi-neutral set of single-fluid equations and resistive MHD conductivity,
\begin{eqnarray}
\Partby t \rho + \Div \Brackets{\rho \vec{u}}=0 \\
\rho \Deby{t}\vec{u}=-\Div \dyad{P}+ \vec{j}\times \vec{B} \\
\frac{\partial}{\partial t} \dyad{P}= -\dyad{P} \cdot \nabla \vec{u} - (\Div \vec{u})\dyad{P}-\vec{u}  \Leftnabla \cdot \dyad{P}-\vec{u}\cdot\nabla\dyad{P} -\Div \dyad{Q} \label{enTensor}\\
\vec{E}+\vec{u}\times\vec{B}=\frac{\vec{j}}{\sigma} \label{conductivity}\\
\Curl \vec{B}=\vec{j} \\
\Div \vec{B}=0 \\
\Curl \vec{E} = -\frac{\partial}{\partial t}\vec{B} 
\end{eqnarray}
where we have denoted the derivative to the left as $\Leftnabla$ in the evolution equation for the stress tensor $\dyad{P}$. These fluid equations may be found in the usual manner by successive moments of the Boltzmann equation, and have been rigorously re-derived in the collisionless limit by Ramos \cite{Ramos2005}. In the energy equation (\Equation{enTensor}), we have neglected resistive dissipation of pressure by removing terms resembling $\Brackets{\vec{E}+\vec{u}\times\vec{B}}\vec{j}$, and dropped terms similar to $\dyad{P}\times \vec{B}$ in anticipation of the form of the pressure tensor. \Equation{enTensor} contains a third rank tensor $\dyad{Q}$ which describes heat flow. 

For the normal mode analysis of this paper  we have assumed time dependent oscillations of the form $\exp{\Brackets{\lambda t}}$ where $\lambda=-i\omega$, and the system of equations is linearised around a non-flowing anisotropic equilibrium as 
\begin{eqnarray}
\vec{B}=\vec{B}_0+\widetilde{\vec{B}} \label{lin1}\\
\vec{E}=\widetilde{\vec{E}}=-\lambda \vec{A} \\
\vec{u}=\widetilde{\vec{V}} \\
\widetilde{\vec{B}}=\Curl \vec{A} \label{lin4}\\
\dyad{P}=\dyad{P}_0+\widetilde{\dyad{P}}\\
\Div \dyad{P}_0=\Curl \vec{B}_0 \times \vec{B}_0 \label{equilibriumVec}\\
\dyad{Q}=\widetilde{\dyad{Q}}
\end{eqnarray}
which leads to the linear unclosed set of equations
\begin{eqnarray}
\lambda \rho_0 \widetilde{\vec{V}} = -\Div \widetilde{\dyad{P} }+ \vec{H} \\
\lambda \widetilde{\dyad{P}} = -\dyad{P}_0 \cdot \nabla \widetilde{\vec{V}}  - (\Div \widetilde{\vec{V}} )\dyad{P}_0-\widetilde{\vec{V}} \Leftnabla \cdot \dyad{P}_0-\widetilde{\vec{V}} \cdot\nabla \dyad{P}_0 -\Div \widetilde{Q}\\
\lambda \vec{A} = - \vec{B}_0\times \widetilde{\vec{V}} - \frac{1}{\sigma} \nabla\times\nabla\times\vec{A} \\
\vec{H}\equiv (\nabla \times \vec{B}_0)\times (\nabla \times \vec{A})-\vec{B}_0\times(\nabla \times \nabla \times \vec{A})
\end{eqnarray}
where $\vec{H}$ may be thought of as a perturbed $\vec{j}\times\vec{B}$ force. We are now in a position to seek an acceptable fluid closure. In this work, we seek to generalise the well-understood MHD fluid model, thus we need a linearised energy equation resembling
\begin{eqnarray}
\lambda \tilde{p} = -\gamma p \Div \widetilde{\vec{V}} - \widetilde{\vec{V}} \cdot \nabla p
\end{eqnarray}
where $\gamma=\frac{5}{3}$, in the isotropic limit. To reproduce the physics of MHD, we must also neglect finite Larmor radius (FLR) effects in our fluid treatment. It is well known  \cite{Chew1956} (and easily verified using drift currents to lowest order in Larmor radius), that the pressure tensor in collisionless plasma may be written as a diagonal matrix
\begin{eqnarray}
\dyad{P}_0=\pperp\dyad{I}+\Delta\vec{B}_0\vec{B}_0, \Delta\equiv\frac{\ppar-\pperp}{\Bsq} \label{CGLP}
\end{eqnarray}
Setting $Q=0$ and $\sigma \to \infty$ reproduces the CGL double-adiabatic closure $\frac{d}{dt}\Brackets{\frac{\pperp}{\rho B}}=0,\frac{d}{dt}\Brackets{\frac{\ppar B^2}{\rho^3}}=0$ \cite{Chew1956}. It may be readily seen that the simple CGL closure does not reduce to the MHD expression $\frac{d}{dt}\Brackets{\frac{p }{\rho^{5/3}}}=0$ in the isotropic limit. Fundamentally, the physical difference between models is that the parallel and perpendicular pressures in MHD are combined and indistinguishable when doing adiabatic work whereas in CGL they are mutually insulated and do adiabatic work independently. Thus to correctly generalise MHD to anisotropic equilibria, we require a `single-adiabatic' closure that somehow treats these pressures together.
 
With some trial and error, we learn that a single-adiabatic theory can be obtained by defining an isotropic perturbed pressure $\tilde{p}$ and assuming zero net heat flow. All anisotropic components to the pressure forces are ignored in order to reproduce only the MHD perturbed forces. To close the equations, our formal thermal assumptions of the model are 
\begin{eqnarray}
\widetilde{\dyad{P}}\to \tilde{p} \dyad{I}+ \widetilde{\dyad{\pi}}\\
\text{Tr}~\Div \widetilde Q \to 0 \\
\text{Tr}~\dyad{\widetilde{\dyad{\pi}}} \to 0
\end{eqnarray}
which, when taking the trace of our energy equation, lead immediately to the linearised model
\begin{eqnarray}
\lambda \rho_0 \widetilde{\vec{V}} = -\nabla \tilde p + \vec{H} \label{momentum}\\
\lambda \tilde{p} = -\Brackets{\frac{4}{3} \pperp + \frac{1}{3} \ppar} \Div \widetilde{\vec{V}} -\frac{2}{3}\Delta \vec{B}_0\vec{B}_0 \colon \nabla \widetilde{\vec{V}} \nonumber \\- \frac{1}{3}\widetilde{\vec{V}} \cdot \nabla(2 \pperp +\ppar) \label{EulerPressure} \\
\lambda \vec{A} = - \vec{B}_0\times \widetilde{\vec{V}} - \frac{1}{\sigma} \nabla\times\nabla\times\vec{A} 
\end{eqnarray}
which reduces to MHD in the isotropic limit as required. 
This heat equation containing the constructed scalar pressure $\tilde{p}$ exhibits a viscous term $\frac{2}{3}\Delta \vec{B}_0\vec{B}_0 \colon \nabla \widetilde{\vec{V}}$ which implies that the differential operator of the eigenvalue problem will be asymmetric and perhaps non-Hermitian, although we have not yet encountered complex eigenvalues in any solutions.

In the next section, we explore the physical interpretation and thermodynamics of this approach and closure.
\subsection{Interpretation of the single and double adiabatic closures}
The fluid stress tensor $\dyad{P}$ may always be decomposed into diagonal and non-diagonal contributions
\begin{eqnarray}
\dyad{P}=\left(
\begin{array}{ccc}
 p_x & 0  & 0  \\
 0 & p_y  & 0  \\
 0 & 0  &  p_z 
\end{array}
\right)+\left(
\begin{array}{ccc}
 0&  \tau_{xy}  & \tau_{xz}  \\
 \tau_{xy} & 0  & \tau_{yz}  \\
 \tau_{xz} & \tau_{yz} &  0 
\end{array}
\right)
\end{eqnarray}
The diagonal contribution represents thermal momentum times a thermal velocity in the same direction, and therefore represents the pressure forces along each degree of freedom, which are anisotropic in general. Conversely, the off-diagonal elements represent thermal momentum transported in orthogonal directions, and so these elements represent shear viscous forces. A further decomposition is possible which defines a scalar part of the pressure and a `parallel viscosity' correction to the scalar expression but leaving the shear viscous off-diagonal portion unchanged
\begin{eqnarray}
\fl \dyad{P}=\left(
\begin{array}{ccc}
 \frac{1}{3}\Brackets{p_x+p_y+p_z} & 0  & 0  \\
 0 &  \frac{1}{3}\Brackets{p_x+p_y+p_z}   & 0  \\
 0 & 0  &   \frac{1}{3}\Brackets{p_x+p_y+p_z} 
\end{array}
\right) + \nonumber \\
\Brackets{ 
\begin{array}{ccc}
 \frac{1}{3}\Brackets{2p_x-p_y-p_z} & 0  & 0  \\
 0 &  \frac{1}{3}\Brackets{-p_x+2p_y-p_z}   & 0  \\
 0 & 0  &   \frac{1}{3}\Brackets{-p_x-p_y+2p_z} 
\end{array}} + \nonumber \\
\left(
\begin{array}{ccc}
 0&  \tau_{xy}  & \tau_{xz}  \\
 \tau_{xy} & 0  & \tau_{yz}  \\
 \tau_{xz} & \tau_{yz} &  0 
\end{array}
\right)
\end{eqnarray}
Comparing to the classical transport theory \cite{Braginskii1965,Marshall1957}, this decomposition is written $\dyad{P}=p\dyad{I}+\dyad{\pi}$ where $p$ is due to the Maxwellian part of the distribution function, and $\dyad{\pi}$ is the non-thermal and traceless remainder, also known as the `dissipative stress tensor'. Conversely, in the collisionless plasma theory, it is well known \cite{Kaufman1960} that the shear off-diagonal terms are zero to first order in Larmor radius when the characteristic time of the phenomena is slow compared to the cyclotron motion. We henceforth drop the gyro-viscous FLR terms. Applying a perturbation to the pressures, we assume symmetry in  $x$ and $y$ for $z$ parallel to the equilibrium field.
\begin{eqnarray}
 \Brackets{\dyad{P}+\widetilde{\dyad{P}}}_{xx,yy}= \frac{1}{3}\Brackets{2p_\perp+2\tilde{p}_\perp+p_\parallel+\tilde{p}_\parallel}+ \nonumber \\
\frac{1}{3}\Brackets{p_\perp+\tilde{p}_\perp-p_\parallel-\tilde{p}_\parallel}\\
\Brackets{\dyad{P}+\widetilde{\dyad{P}}}_{zz}= \frac{1}{3}\Brackets{2p_\perp+2\tilde{p}_\perp+p_\parallel+\tilde{p}_\parallel}+ \nonumber \\
\frac{2}{3}\Brackets{-p_\perp-\tilde{p}_\perp+p_\parallel+\tilde{p}_\parallel}
\end{eqnarray}
We may identify the terms $p=\frac{1}{3}\Brackets{2\pperp + \ppar}$, $\pi_\perp=\frac{1}{3}\Brackets{p_\perp-p_\parallel}$,  $\pi_\parallel=\frac{2}{3}\Brackets{\ppar-\pperp}$ and the perturbed versions $\tilde{p}=\frac{1}{3}\Brackets{2\tilde{\pperp} + \tilde{\ppar}}$, $\tilde{\pi_\perp}=\frac{1}{3}\Brackets{\tilde{p_\perp}-\tilde{p_\parallel}}$,  $\tilde{\pi_\parallel}=\frac{2}{3}\Brackets{\tilde{\ppar}-\tilde{\pperp}}$. 
We now write the linearised energy equation for each degree of freedom
\begin{eqnarray}
\lambda \widetilde{\dyad{P}}_{xx}=-2\Brackets{\Brackets{p+\pi_\perp} \partial_x}\widetilde{V_{x}}\nonumber\\
-\Brackets{p+\pi_\perp}\Div \widetilde{\vec{V}}-\widetilde{\vec{V}}\cdot \nabla \Brackets{p+\pi_\perp}-dQ_{xx}\\
\lambda \widetilde{\dyad{P}}_{yy}=-2\Brackets{\Brackets{p+\pi_\perp} \partial_y}\widetilde{V_{y}}\nonumber\\
-\Brackets{p+\pi_\perp}\Div \widetilde{\vec{V}}-\widetilde{\vec{V}}\cdot \nabla \Brackets{p+\pi_\perp}-dQ_{yy}  \\
\lambda \widetilde{\dyad{P}}_{zz}=-2\Brackets{\Brackets{p+\pi_\parallel} \partial_z}\widetilde{V_{z}}\nonumber\\
-\Brackets{p+\pi_\parallel}\Div \widetilde{\vec{V}}-\widetilde{\vec{V}}\cdot \nabla \Brackets{p+\pi_\parallel}-dQ_{zz}
\end{eqnarray}
simplifying to
\begin{eqnarray}
\lambda (\tilde{p} +\tilde{\pi}_\perp)=\Brackets{p+\pi_\perp}\partial_z\widetilde{V_z}-2\Brackets{p+\pi_\perp}\Div \widetilde{\vec{V}}\nonumber \\
-\widetilde{\vec{V}}\cdot \nabla \Brackets{p+\pi_\perp} -dQ_\perp\label{CGL1}\\
\lambda (\tilde{p} + \tilde{\pi}_\parallel)=-2\Brackets{p+\pi_\parallel} \partial_z\widetilde{V_{z}}-\Brackets{p+\pi_\parallel}\Div \widetilde{\vec{V}} \nonumber\\
-\widetilde{\vec{V}}\cdot \nabla \Brackets{p+\pi_\parallel} -dQ_\parallel \label{CGL2}
\end{eqnarray}
which is the double-adiabatic CGL fluid closure if $dQ_\perp$ and $dQ_\parallel$ are zero. 

To recover some hydrodynamic intuition and return to an MHD-like model, we turn our attention to the behaviour of the isotropic `average' pressure $\tilde{p}$ which underlies the CGL model before anisotropic perturbed forces are included. We simply eliminate $\tilde{\pi}_\perp$ and $\tilde{\pi}_\parallel$ from our equations using the identity $2\tilde{\pi}_\perp+\tilde{\pi}_\parallel=0$; adding twice \Equation{CGL1} to \Equation{CGL2}  with zero heat flow we obtain 
\begin{eqnarray}
\lambda \tilde{p} =\frac{2}{3}\Brackets{\pi_{\perp }-\pi_{\parallel }}\partial_z\widetilde{V}_z-\Brackets{\gamma p+\frac{\pi_{\perp}+\pi_{\parallel}}{3}}\Div \widetilde{\vec{V}} -\widetilde{\vec{V}}\cdot \nabla p \label{SA}
\end{eqnarray}
which is our scalar \Equation{EulerPressure} derived earlier. In the next section, we investigate the thermodynamics of this model.
\subsection{Thermodynamics}
We consider the thermodynamics of a Lagrangian fluid element of length $z$, cross-sectional area $A$ and volume $V=zA$ which obeys the single-adiabatic closure
\begin{eqnarray}
dQ=0\\
d\pperp \equiv dp+d\pi_\perp \\
d\ppar \equiv dp+d\pi_\parallel \\
2d\pi_\perp+ d\pi_\parallel \equiv 0
\end{eqnarray}
where the fluid element is sufficiently small to be homogenous, and all changes are quasi-static in the classical thermodynamics sense. The condition $dQ=dQ_\parallel+dQ_\perp=0$ is a weaker assumption than the CGL condition $dQ_\perp=dQ_\parallel=0$. This is a luxury afforded to us because of our neglect of detailed behaviour of the parallel and perpendicular motion, which requires additional assumptions on $d\pi_\perp$ and $d\pi_\parallel$ to close the equations. In the CGL model, the condition $dQ_\perp=dQ_\parallel=0$ provides that additional closure. Conversely, in the SA theory, we make no such statement, leaving open the possibility of non-adiabatic internal degrees of freedom.
\subsubsection{First law}
Treating the parallel and perpendicular degrees of freedom as separate energy reservoirs doing work whilst exchanging heat between them leads us to the relations in terms of energy $U$, heat $Q$ and work $W$
\begin{eqnarray}
U_\parallel=\frac{p_\parallel}{2} V \\
U_\perp=p_\perp V \\
dU=dQ-dW=\Brackets{dQ_\parallel+dQ_\perp}-\Brackets{dW_\parallel +dW_\perp} \\
dU_\parallel=dQ_\parallel-dW_\parallel \\
dU_\perp=dQ_\perp-dW_\perp \\
dW_\parallel =\ppar A dz \\
dW_\perp =\pperp z dA 
\end{eqnarray}
differentiation and some algebraic manipulation of the above relations gives the implied perturbed pressure 
\begin{eqnarray}
dp=-\Brackets{\frac{4}{3}\pperp +\frac{1}{3}\ppar}\frac{dA}{A}-\Brackets{\ppar+\frac{2}{3}\pperp}\frac{dz}{z} \label{dpthermo}
\end{eqnarray}
which is derived from the first law  applied to the energy in each separate degree of freedom as well as the total energy.  To remove any further doubt that this energy conserving system corresponds exactly to our closure, we reformulate our Eulerian perturbation \Equation{EulerPressure} instead as a Lagrangian perturbation \cite{Frieman1960} of the pressure
\begin{eqnarray}
\lambda \vec{\xi}=\widetilde{\vec{V}} \\
\dyad{P}(\vec{r}^0+\vec{\xi})=\nonumber \\
\dyad{P(\vec{r}^0)}-\vec{\xi}\Brackets{\Leftnabla^0 \cdot \dyad{P(\vec{r}^0)}}-\Brackets{\dyad{P(\vec{r}^0)} \cdot \nabla^0}\vec{\xi}-\dyad{P(\vec{r}^0)}\Brackets{\nabla^0 \cdot \vec{\xi}}\\
\nabla\approx\nabla^0-\nabla^0\xi\cdot\nabla^0 \\
\lambda \hat{p} = -\Brackets{\frac{4}{3}\pperp+\frac{1}{3} \ppar } \nabla^0\cdot \widetilde{\vec V} -\frac{2}{3} \Delta  B_{0i} B_0^j \nabla^0_j \widetilde{V}^i   ~(*)
\end{eqnarray}
integrating over the fluid volume element $xyz=Az$ and identifying $\vec{\xi}\equiv (dx,dy,dz)$ and $\vec{B_0}=B_0 \hat{z}$ gives immediately \Equation{dpthermo} and \Equation{SA}. Thus, the SA force operator conserves total energy.  It simply doesn't cover nor restrict the detailed energy conservation or motion of the parallel and perpendicular components.
\subsubsection{Second law}
We now show that the heat flows $dQ,dQ_\perp$ and $dQ_\parallel$ satisfy the second law of thermodynamics $dS\ge 0$. Without loss of generality, we define Bi-Maxwellain temperatures in terms of the equilibrium density and pressures $\pperp\equiv\rho T_\perp$ and $\ppar\equiv\rho T_\parallel$. The respective relations for the entropy changes are 
\begin{eqnarray}
dS_\perp = \frac{dQ_\perp}{T_\perp} \\
dS_\parallel = \frac{dQ_\parallel}{T_\parallel} 
\end{eqnarray}
Using $dQ_\perp+dQ_\perp=dQ=0$ we immediately obtain
\begin{eqnarray}
\frac{dQ_\perp}{T_\perp}\Brackets{1-\frac{T_\perp}{T_\parallel}} = dS_\perp +dS_\parallel 
\end{eqnarray}
Consider first the case $T_\perp > T_\parallel $; since $dQ=0$, $dQ_\perp$ must be negative because heat must flow from perpendicular to parallel (hot to cold), this implies $dS=dS_\perp + dS_\parallel > 0$. Conversely, when $T_\perp < T_\parallel $, $dQ_\perp$ is positive which still implies $dS=dS_\perp + dS_\parallel > 0$. We may now relax the Bi-Maxwellian condition for each degree of freedom and include further entropy increases due to relaxation, leaving the result unchanged.

This places a clear equilibrium-determined `must not oscillate' restriction on $dQ_\perp$ for any valid fluid theory, since the equilibrium doesn't oscillate by definition. Checking the behaviour of sound waves $dA=0$ under the SA model we obtain
\begin{eqnarray}
\frac{dQ_\parallel}{V}=\frac{dp}{2}+\frac{d\pi_\parallel}{2}+2\ppar\frac{dz}{z} \\
\frac{dQ_\perp}{V}=dp+d\pi_\perp+\pperp\frac{dz}{z} \\
dp=-\ppar \frac{dz}{z}-\frac{2\pperp}{3}\frac{dz}{z}
\end{eqnarray}
This shows that $dp$ may admit strictly oscillatory sound wave solutions without violating the second law requirement that $dQ_\perp$ must not change sign. Substituting this oscillatory pressure into the previous two equations for the heat flow causes no contradiction, since the residual anisotropic parts $d\pi_\parallel$ and $d\pi_\perp$ are left unconstrained. In passing, we mention that for CGL, these terms may also oscillate along with $p$, but for finite heat flow, they must not without some other internal relaxation process to increase the entropy. This behaviour is analogous to the classical thermodynamics result for the modification of ideal gas sound waves due to the relaxation of internal degrees of freedom.

Thus, the SA model satisfies the second law of thermodynamics because $dQ$ is zero, and no further restrictions exist on the internal heat flow. 
\section{Equations for a MISHKA-like numerical code}
In this section, we deliberately follow closely the structure and notation found in \cite{ISI:A1997YB84300003} to re-derive an anisotropic set of equations for the linear normal-mode MHD solver MISHKA, briefly restating, but not dwelling on, unchanged results. For convenient reference, we will denote changed equations with an asterisk `$(*)$'. The presented perturbed pressure terms, for the sake of brevity, will be for the novel single-adiabatic model only, although we anticipate that implementation in MISHKA will also include the CGL expressions for comparison.
\subsection{Equilibrium}
We proceed in the straight field line coordinate system $\vec r=\vec r(s,\upvartheta,\phi),$ where $s\equiv \sqrt{\frac{\psi}{\psi_{\text{edge}}}}$ and the Jacobian is chosen such that the angle $\phi$ is the ordinary toroidal angle in cylindrical coordinates $(R,Z,\phi)$ , then the contravariant components of the equilibrium field, and metric tensor, are
\begin{eqnarray}
B_0^1=0  \\
\frac{B_0^3}{B_0^2} \equiv q(s) \\
g^{33}\equiv \nabla \phi \cdot \nabla \phi = \frac{1}{R^2} \\
g_{33}\equiv  \partial_\phi \vec r \cdot \partial_\phi \vec r = R^2 
\end{eqnarray}
the equilibrium magnetic field in covariant form may be written
\begin{eqnarray}
\vec B_0 = \nabla \phi \times \nabla \psi(s) + F(s,\upvartheta) \nabla \phi ~(*) \\
F(s,\upvartheta) =F_\Delta(s)/(1-\Delta(s,\upvartheta))  ~(*) \\
B_0^2=\psi'(s)/J\equiv f/J 
\end{eqnarray}
where the toroidal field is a flux function for the special case of isotropic equilibrium  $R B_\phi \equiv F(s,\upvartheta) =F_\Delta(s)$ \cite{Dobrott1970}.  Summarising, the Jacobian, covariant and contravariant components of the equilibrium field are:
\begin{eqnarray}
J=\frac{qfR^2}{F} \\
B_0^1=0 \label{B0U1}\\
B_0^2=\frac{F}{qR^2} \label{B0U2}\\
B_0^3=\frac{F}{R^2} \label{B0U3}\\
B_{01}=g_{12}\frac{F}{qR^2} \label{B0D1}\\
B_{02}=g_{22}\frac{F}{qR^2} \label{B0D2}\\
B_{03} = F \label{B0D3}
\end{eqnarray}
Henceforth we take as understood the standard notation from the tensor calculus, such as superscript and subscript denoting contravariant and covariant components of vectors, as well as the metric tensor $g_{ij}$ and Christoffel symbols $\Gamma_{ij}^{k}$. Taking the curl of the equilibrium field gives the equilibrium current density. Doing so with particular care to include the anisotropic $\upvartheta$ dependence in $F$, the components are
\begin{eqnarray}
j_0^1=\frac{1}{J}\Partby{\upvartheta}F~(*) \label{j01}\\
j_0^2=-\frac{1}{J}\Partby{s} F\\
j_0^3=\frac{1}{J}\SBrackets{\Partby s \Brackets{g_{22} \frac{F}{qR^2}}-\frac{1}{q}\Partby \upvartheta \Brackets{g_{12}\frac{F}{R^2}}} ~(*) \label{j03}
\end{eqnarray}
We note in particular that current no longer lies in flux surfaces for anisotropic equilibrium. For completeness here, we include the integrated fluxes and currents given by
\begin{eqnarray}
\chi'(s)=-2\pi f \\
\Phi'(s)=2 \pi f q \\
I_\upvartheta(s,\upvartheta)=2 \pi F(s,\upvartheta) \\
I_\phi(s)=\frac{2\pi}{q}\left <\frac{g_{22}F}{R^2} \right >_\upvartheta ~(*)
\end{eqnarray}
To obtain a Grad-Shafranov type equation, we take the radial component of the pressure
\begin{eqnarray}
\SBrackets{\Div\dyad{P_0}}_1=\Partby s \pperp + \frac{1}{J}\partial_i\Brackets{J\Delta B_0^i g_{21}B_0^2}-\Delta B_0^i B_{0j}\Gamma_{i1}^j
\end{eqnarray}
and, using \EqnsMany{\ref{j01}}{\ref{j03}}, equate it to the radial component of the $\vec{j}_0\times \vec{B}_0$ force in \Equation{equilibriumVec} giving
\begin{eqnarray}
\frac{F}{R^2}\Partby s F + \frac{F}{qR^2}\SBrackets{\Partby s \Brackets{\frac{F}{qR^2}g_{22}}-\frac{1}{q}\Partby \upvartheta \Brackets{\frac{F}{R^2}g_{12}}}= \nonumber \\ -\Partby s \pperp -\frac{1}{J}\partial_i\Brackets{J\Delta B_0^i g_{12}\frac{F}{qR^2}}+\Delta B_0^i B_{0j}\Gamma_{i1}^j ~(*) \label{GS}
\end{eqnarray}
\Equation{GS} is a form of the equilibrium condition which needs to be solved for numerical stability analysis with high precision in a code such as HELENA to obtain the straight-field line geometry inputs for a linear stability code such as MISHKA.  The isotropic form \cite{ISI:A1997YB84300003} is regained in the isotropic limit $\Delta \to 0$ with $F$ and $\pperp$ becoming flux functions. The necessary modifications to HELENA to incorporate \Equation{GS} have been performed by Qu et al. \cite{Qu2014} who also analyse the magnitude of the anisotropic effects on equilibrium. We will leave further discussion on the topic of tokamak equilibrium here and proceed to the linear stability equations in toroidal geometry
\subsection{Projections of perturbed quantities and conductivity equation}
In addition to the simplification introduced by the straight field line coordinate system, the MHD stability problem is simplified when the perturbed vector fields $\vec{A} ,\vec{\widetilde{V}}$ are projected along normal, binormal and parallel directions.
In the MISHKA codes, the projections chosen are
\begin{eqnarray}
\hat A_2 \equiv \SBrackets{\vec A \times \vec B_0}^1/\Bsq \\
\hat A_3 \equiv \vec A \cdot \vec B_0 / \Bsq \\
\hat V^2 \equiv \SBrackets{\widetilde{ \vec V} \times \vec B_0}_1 \\
\hat V^3 \equiv \widetilde{ \vec V} \cdot \vec B_0 / \Bsq
\end{eqnarray}
which relate to the poloidal and toroidal components using \EqnsMany{\ref{B0U1}}{\ref{B0D3}}
\begin{eqnarray}
A_2= fq \hat A_2 + g_{22}\frac{F}{qR^2}\hat A_3\\
A_3=-f \hat A_2 + F \hat A_3 \\
\widetilde V^2 = \frac{F^2}{qR^2\Bsq}\Brackets{-\frac{g_{21}}{qR^2}\widetilde V^1+\frac{1}{f}\hat V^2} +\frac{F}{qR^2}\hat V^3 \label{VU1}\\
\widetilde V^3 =-\frac{F^2}{qR^4\Bsq}\Brackets{g_{12}\widetilde V^1 + \frac{g_{22}}{qf}\hat V^2} + \frac{F}{R^2}\hat V^3 \label{VU2}
\end{eqnarray}
Taking the curl of $\vec{A}$ gives the perturbed magnetic field
\begin{eqnarray}
\widetilde B^1 = \frac{1}{J} \SBrackets{-f \Brackets{\Partby \upvartheta + q \Partby \phi}\hat A_2+\Brackets{\Partby \upvartheta - \frac{g_{22}}{qR^2}\Partby \phi}F\hat A_3} ~(*) \label{B11}\\
\widetilde B^2 = \frac{1}{J} \SBrackets{\Partby \phi A_1+\Partby s \Brackets{f \hat A_2}- \Partby s \Brackets{F \hat A_3}} \\
\widetilde B^3 =  \frac{1}{J} \SBrackets{-\Partby \upvartheta A_1+\Partby s \Brackets{fq\hat A_2} + \Partby s \Brackets{\frac{g_{22}F}{qR^2}\hat A_3}}\label{B13}
\end{eqnarray}
where the perturbed radial magnetic field has been modified by the poloidal dependence of $F$.  We may now use these projections for current, magnetic field and fluid velocity to obtain the equations of motion from the linearised momentum, conductivity and energy equations. The closed system in MISHKA requires equations in terms of equilibrium quantities and the seven unknowns $\widetilde{V}^1, \hat{V}^2, \hat{V}^3,  A_1, \hat{A}_2, \hat{A}_3$ and $\widetilde{p}$.

We first deal quickly with the Ohm's law \Equation{conductivity} and its equations relating $\vec{A}$ and $\widetilde{\vec{V}}$ because they are dependant on Maxwell's equations and the linearisation \EqnsMany{\ref{lin1}}{\ref{lin4}} but not on the pressure. The unchanged isotropic resistive MHD results are \cite{ISI:A1997YB84300003}
\begin{eqnarray}
\lambda A_1=\hat{V}^2 - \frac{1}{\Bsq \sigma}\Brackets{\Curl \Curl \vec{A}}_1  \label{A1}\\
\lambda \hat{A}_2=-\hat{V}^1 - \frac{1}{\Bsq \sigma}\Brackets{\Brackets{\Curl \Curl \vec{A}}\times\vec{B}_0}^1\label{A2}\\
\lambda \hat{A}_3=\frac{1}{\Bsq \sigma} \vec{B}_0\cdot \Brackets{\Curl \Curl \vec{A}} \label{A3}
\end{eqnarray}
where the parallel electric field proportional to $\hat{A}_3$ is zero for ideal MHD. The parallel magnetic perturbations due to the resistive terms in \EqnsMany{\ref{A1}}{\ref{A2}} are negligible for the shear \Alfven~wave.
\subsection{Momentum equation}
Seeking a set of expressions for the velocities $\widetilde{V}^1, \hat{V}^2, \hat{V}^3$ in terms of the remaining MISHKA variables, we turn our attention to the linearised momentum \Equation{momentum}.
The radial and poloidal components of the momentum equation
\begin{eqnarray}
\lambda \rho_0 \widetilde V_1 = -\Partby s \tilde p + H_1 \label{momentum1}\label{V1}\\
\lambda \rho_0 \widetilde V_2 = -\Partby \upvartheta \tilde p + H_2 \label{momentum2} \label{V2}
\end{eqnarray}
contain velocity and perturbed $\vec{j}\times\vec{B}$ terms which are dealt with first.
\EqnsMany{\ref{B0U1}}{\ref{B0U3}}, \EqnsMany{\ref{j01}}{\ref{j03}} and \EqnsMany{\ref{B11}}{\ref{B13}} are combined to produce the perturbed $\vec{j}\times \vec{B}$ terms giving
\begin{eqnarray}
H_1=\alpha_1 - \frac{F}{R^2}\Partby s \Brackets{JG\widetilde B^3} \\
H_2=\alpha_2 - \frac{F}{R^2}\Partby \upvartheta \Brackets{JG\widetilde B^3} ~(*) \\
\alpha_1\equiv J\Brackets{j_0^2\widetilde{B}^3-j_0^3\widetilde{B}^2} - \frac{F}{qR^2}\Partby s \Brackets{JM\widetilde{B}^1 + JN\widetilde{B}^2 }  \nonumber \\
+\frac{F}{qR^2}\Brackets{\Partby \upvartheta + q \Partby \phi}\Brackets{JL\widetilde B^1 +JM\widetilde B^2 } \\
\alpha_2\equiv J\Brackets{j_0^3\widetilde{B}^1-j_0^1\widetilde{B}^3} + \frac{F}{R^2}\Partby \phi \Brackets{JM\widetilde{B}^1 + JN\widetilde{B}^2 } (*)\\
L\equiv \frac{g_{11}}{J}, M\equiv \frac{g_{12}}{J}, N\equiv \frac{g_{22}}{J}, G\equiv \frac{g_{33}}{J}=\frac{F}{fq} 
\end{eqnarray}
and the covariant velocity terms on the left hand side of \Eqns{\ref{momentum1}}{\ref{momentum2}}, using \Eqns{\ref{VU1}}{\ref{VU2}}, are 
\begin{eqnarray}
\widetilde{V}_1 = \Brackets{g_{11}-\frac{g_{21}^2F^2}{q^2R^4\Bsq}}\widetilde{V}^1 + \frac{g_{12}F}{qR^2}\Brackets{\frac{F}{f\Bsq}\hat V^2 +\hat V^3}\\
\widetilde {V}_2=\frac{g_{12}F^2}{R^2\Bsq} \widetilde{V}^1 + \frac{g_{22}F}{qR^2}\Brackets{\frac{F}{f\Bsq}\hat V^2 + \hat V^3}
\end{eqnarray}
The remaining momentum equation is the component of \Equation{momentum} parallel to the field obtained by dot product with $\vec{B}_0$ and inclusion of the equilibrium \Equation{GS}, giving
\begin{eqnarray}
\lambda \rho_0 \Bsq \hat V^3=-\frac{F}{qR^2}\Brackets{\Partby \upvartheta + q \Partby \phi} \tilde p \nonumber \\
- \Brackets{\widetilde B^1 \Partby s + \widetilde B^2 \Partby \upvartheta } \pperp-\Delta \widetilde B^i B_0^j \nabla_j B_{0i} \nonumber \\ -\Brackets{\widetilde B^1g_{12}\frac{F}{qR^2} +\widetilde B^2g_{22}\frac{F}{qR^2}+\widetilde B^3 F} \frac{F}{qR^2}\Partby \upvartheta\Delta ~(*) \label{V3}
\end{eqnarray}
which is a far more complex expression for parallel momentum than the MHD version, due to the  difficulty of taking a derivative of anisotropic equilibrium pressure in the direction of the perturbed field. Pressure and magnetic surfaces no longer coincide and so changes in pressure cannot be represented by a radial derivative alone. Furthermore, owing to the average gyro-motion of a magnetised plasma, the pressure tensor is oriented with the equilibrium field direction through the $\Delta \vec{B}_0 \vec{B}_0$ part of \Equation{CGLP}, so a derivative of pressure in a direction tangent to a field line must include changes due to curvature, as expressed with $\Delta \widetilde B^i B_0^j \nabla_j B_{0i}$. Thus, the introduction of anisotropy means that curvature terms must be computed explicitly from the metric tensor and included in the numerical treatment. 
\subsection{Energy equation}
Having obtained six equations (\Equation{A1}, \Equation{A2}, \Equation{A3}, \Equation{V1},\Equation{V2} and \Equation{V3}) relating the seven components of $\vec{A}$, $\widetilde{\vec {V} }$ and $\widetilde{p}$, we close the system of equations with the energy equation
\begin{eqnarray}
\lambda \tilde{p} = -\Brackets{\frac{4}{3}\pperp+\frac{1}{3} \ppar } \Div \widetilde{\vec V} -\frac{2}{3} \Delta  B_{0i} B_0^j \nabla_j \widetilde{ V}^i  \nonumber \\ -\widetilde{\vec V} \cdot \nabla\Brackets{ \frac{2}{3} \pperp +  \frac{1}{3}\ppar} ~(*) \label{ptilde}
\end{eqnarray}
and the derivative terms in the MISHKA variables become
\begin{eqnarray}
J\Div \widetilde{\vec V} = \nonumber \\ \Brackets{\Partby s J \widetilde{ V}^1}-\frac{f}{q}\Brackets{\Partby \upvartheta + q \Partby \phi}\frac{g_{12}F}{\Bsq R^2}\widetilde{V}^1 \nonumber \\
+\Brackets{\Partby \upvartheta - \frac{g_{22}}{qR^2}\Partby \phi}\frac{F}{\Bsq}\hat V^2 + f\Brackets{\Partby \upvartheta + q \Partby \phi}\hat V^3 ~(*)  \nonumber \\
\widetilde{\vec V} \cdot \nabla\Brackets{ \frac{2}{3} \pperp +  \frac{1}{3}\ppar} =\nonumber \\ \widetilde{V}^1 \Partby s \Brackets{ \frac{2}{3} \pperp +  \frac{1}{3}\ppar} \nonumber \\ + \SBrackets{\frac{F^2}{qR^2\Bsq}\Brackets{\frac{\hat V^2}{f}-\frac{g_{12}\widetilde{V}^1}{qR^2}}+\frac{F}{qR^2}\hat V^3}\Partby \upvartheta \Brackets{ \frac{2}{3} \pperp +  \frac{1}{3}\ppar} ~(*)
\end{eqnarray}
Again we note here that the angular dependence of equilibrium pressure and curvature complicates the expressions, so much so that it is not worthwhile to change from Eulerian pressure $\widetilde p$ to Lagrangian pressure $\hat{p}$ as done in MISHKA.

The viscous double product  $\frac{2}{3} \Delta  B_{0i} B_0^j \nabla_j \widetilde{ V}^i$ may be evaluated by tedious inclusion of all Christoffel symbols computed for the equilibrium straight field line coordinates, however some simplification is possible and is discussed in the next section. 
\subsection{Simplification of curvature terms}
The curvature term in \Equation{ptilde} is simplified considerably when expressed in terms of the field curvature $\vec{\kappa} \equiv  \hat{\vec b}\cdot \nabla \hat{\vec b}$ and the orthonormal basis 
\begin{eqnarray}
\hat{\vec{n}}\equiv \frac{\nabla s}{\Norm{\nabla s}},& \hat{\vec \perp} \equiv \hat{\vec b} \times \hat{\vec n}, & \hat{\vec b} \equiv \frac{\vec B_0}{\Norm{\vec B_0}}\end{eqnarray}
which represent the normal, binormal and parallel directions to the flux surface and field line respectively.  Realising that $\bunit\bunit:\nabla\nunit=-\nunit\bunit:\nabla\bunit=-\vec{\kappa}\cdot \nunit$, and similarly for $\hat{\vec \perp}$ instead of $\nunit$, allows a deconstruction of the curvature in terms of normal curvature $\vec{\kappa}\cdot \nunit \equiv \kappa_n$ and geodesic curvature $\vec{\kappa}\cdot  \hat{\vec \perp} \equiv \kappa_g$ (`\emph{ad more geometrico}' in \cite{Goedbloed2010}). Furthermore, the chosen contravariant velocity variables for MISHKA are almost the suitable form for that orthonormal basis, with projections of velocity along the unit vectors
\begin{eqnarray}
\nunit = \frac{\Unit{e}^1}{\sqrt{g^{11}}} \\
\Unit{\perp}=\frac{1}{J\Norm{\vec{B}_0}\sqrt{g^{11}}}\Brackets{B_{03}\Unit{e}_2-B_{02}\Unit{e}_3} \\
\bunit = \frac{1}{\Norm{\vec{B}_0}} \Brackets{B^{2}_0\Unit{e}_2+B^3_0\Unit{e}_3}
\end{eqnarray}
providing the conversion between bases. The final form for the viscous contribution reduces to\begin{eqnarray}
\fl \frac{2}{3}\Brackets{\ppar-\pperp} \times \nonumber \\
\fl \SBrackets{\frac{F}{qR^2}\Brackets{\Partby \upvartheta + q \Partby \phi}\Brackets{\Norm{\vec{B}_0} \hat{V}^3}-\widetilde{V}^1\Brackets{\frac{\kappa_n}{\sqrt{g^{11}}}+\frac{\kappa_g g_{12}F^2 }{fqR^2\Norm{\vec{B}_0}\sqrt{g^{11}}}}-\hat{V}^2\frac{\kappa_g\sqrt{g^{11}}}{\Norm{B}_0}}
\end{eqnarray}
Thus the seven equations \Equation{A1}, \Equation{A2}, \Equation{A3}, \Equation{V1}, \Equation{V2}, \Equation{V3} and \Equation{ptilde} in the seven unknown variables $\widetilde{V}^1, \hat{V}^2, \hat{V}^3, A_1, \hat{A}_2, \hat{A}_3$ and $\widetilde{p}$ constitute a full set of linearised equations of motion for the code MISHKA generalising MHD to anisotropic equilibria. This modification to MISHKA and subsequent numerical calculations is the subject of on-going work and a future publication.
\section{Analytical large aspect ratio and cylinder calculations}
We now finish this paper with an analytical treatment of the large aspect ratio tokamak equilibrium and cylindrical linear modes to give some intuition about the modifications brought by anisotropy to the MHD stability problem, as well as some simple expressions for comparison with experiments in cylindrical symmetry. This will include a modified dispersion relation for the homogeneous and cylindrical continuum.

We consider a shifted circle geometry which is up-down symmetric and with major radius $R(r,\theta) \equiv R_0 + \delta(r) + r \cos \theta$. Then to leading order in inverse aspect ratio $\epsilon \sim r/R$, the straight field line coordinates may be written in orthogonal polar coordinates $(s,\upvartheta) \mapsto (r, \theta)$ and the large aspect ratio metric tensor is given by
\begin{eqnarray} 
g_{ij}=\left(
\begin{array}{ccc}
 1+2\delta' \cos \theta &  -r\delta'\sin\theta  & 0  \\
 -r\delta'\sin\theta &  r^2 &   0\\
 0 &  0 &   R^2
\end{array}
\right) +O(\epsilon^2)
\label{metric}\end{eqnarray}
where setting $\delta \to 0$ gives the zeroth order cylindrical metric.
\subsection{Equilibrium}
Evaluating the equilibrium \Equation{GS} with the large aspect ratio metric gives
\begin{eqnarray}
\frac{B_\phi}{R}\Partby{r}\Brackets{RB_\phi}+\frac{B_\phi}{r} \SBrackets{\Partby{r}\Brackets{rB_\theta} +\Partby{\theta}\Brackets{B_\theta \sin \theta \delta'}} + O(\epsilon^2) = \nonumber \\
-\Partby r \pperp + \Delta\SBrackets{\frac{B_\theta^2}{r}+\frac{B_\phi^2}{R}\Brackets{\cos \theta + \delta'}}+\frac{1}{r}\Partby \theta \Brackets{\Delta B_\theta^2 \sin \theta \delta '} \label{lar}
\end{eqnarray}
where the angular dependence of pressure and the curvature terms are clearly evident. Using the fact that the poloidal flux $\psi$ and $F_\Delta$ are flux functions, we deduce the form for the toroidal and poloidal fields for an anisotropic cylinder to be
\begin{eqnarray}
B_{\phi c}\Brackets{r,\theta}\equiv \Brackets{1-\Delta\Brackets{r,\theta}}\frac{F_\Delta(r)}{R_0} \\
B_{\theta c} \Brackets{r} \equiv \frac{1}{rR_0}\frac{d \psi}{dr}
\end{eqnarray}
Taking the zeroth order of \Equation{lar} gives the cylinder equilibrium equation
\begin{eqnarray}
\Partby {r} \Brackets{\pperp + \frac{1}{2} B_{\phi c}^2} +\frac{B_{\theta c}}{r}\frac{d}{dr}\Brackets{r B_{\theta c}} -\Delta \frac{B_{\theta c}^2}{r} =0 
\end{eqnarray}
noting the partial derivative in the first term. This expression is the well known cylinder force balance condition with the additional term in terms of the poloidal field and anisotropy. This equation may also be quickly checked by direct substitution of the pressure tensor \Equation{CGLP} into $\vec{j}\times\vec{B}=\Div \dyad{P}$ in cylindrical coordinates. The remaining first order terms of \Equation{lar} gives the differential equation for $\delta$
\begin{eqnarray}
\Fullby{r}\Brackets{r B_{\theta c}^2 \delta'}=\frac{r}{R_0}\Brackets{2r\Partby{r}\pperp -B_{\theta c}^2 + 2R_0 \delta ' \frac{\Delta B_{\theta c}^2}{r}-\Delta B_{\phi c}^2}
\end{eqnarray}
where again the perpendicular pressure dominates the force balance, and anisotropic curvature terms proportional to the plasma $\beta$ are the new contributions to the axis shift. In the isotropic limit, the angular dependence of $\pperp$ disappears as do the curvature terms and we are left with the well known equations for the Shafranov shift.
\subsection{Cylindrical wave equation}
We introduce the fluid displacement vector $\widetilde{\vec{V}} \equiv\frac{d}{dt}\vec{\xi}$ to the linear equations of motion
\begin{eqnarray}
-\omega^2 \rho_0 \vec{\xi} = -\nabla \widetilde{p} + \vec{H} \\
\widetilde{p} = -\Brackets{\frac{4}{3}\pperp + \frac{1}{3}\ppar}\Div \vec{\xi}-\frac{2}{3}\Delta\vec{B}_0\vec{B}_0:\nabla\vec{\xi}\nonumber \\
-\vec \xi \cdot \nabla \Brackets{\frac{2}{3}\pperp + \frac{1}{3}\ppar}
\end{eqnarray}
and assume a decomposition in the orthonormal form $\SBrackets{\xi_n(r),\xi_\perp(r),\xi_\parallel(r)}\exp (i\vec{k}\cdot \vec{r}-i\omega t)$ in the normal, binormal and parallel directions. In the cylindrical metric corresponding to the $\epsilon\to 0$ limit in \Equation{metric}, a tedious calculation gives the cylindrical spectral wave equation
\newcommand\Maa{     \Fullby{r}\SBrackets{\frac{ p_f +B_0^2}{r}\Fullby{r}r+ \Lambda}  -B_0^2 \kpar^2-r\Brackets{\frac{B_\theta^2}{r^2}}' }
\newcommand\Mab{     \Fullby{r} \SBrackets{p_f+B_0^2}\kperp - \frac{2kB_\theta B_0}{r}      }
\newcommand\Mac{      \Fullby{r} p_s\kpar  }
\newcommand\Mba{   -\kperp \SBrackets{\frac{p_f+B_0^2}{r}\Fullby{r}r +\Lambda} -\frac{2kB_\theta B_0}{r}  }
\newcommand\Mbb{-\kperp^2\Brackets{ p_f+B_0^2}-\kpar^2B_0^2 }
\newcommand\Mbc{-\kperp\kpar p_s }
\newcommand\Mca{   -\kpar \SBrackets{\frac{p_f}{r}\Fullby{r}r +\Lambda} }
\newcommand\Mcb{-\kpar\kperp p_f}
\newcommand\Mcc{-\kpar^2 p_s}
\begin{eqnarray}
\fl
\left(
\begin{array}{ccc}
 \Maa & \Mab  &  \Mac \\
 \Mba & \Mbb  & \Mbc  \\
 \Mca &  \Mcb &   \Mcc
\end{array}
\right)
\left(
\begin{array}{c}
\xi_n    \\
i\xi_\perp     \\
i\xi_\parallel     
\end{array}
\right)
 \nonumber \\
 =-\rho_0 \omega^2 \left(
\begin{array}{c}
\xi_n    \\
i\xi_\perp     \\
i\xi_\parallel     
\end{array}
\right) \label{cylinderWave}\\
\Lambda\equiv\frac{5}{3}\Delta\frac{B_\theta^2}{r} + \Brackets{\frac{\ppar-\pperp}{3}}' ,
 p_f \equiv \frac{4}{3}\pperp + \frac{1}{3}\ppar ,
 p_s \equiv \frac{2}{3}\pperp + \ppar 
\end{eqnarray}
where we have deliberately re-arranged the equation for side-by-side comparison to the Goedbloed and Poedts MHD text \cite{Goedbloed2004}. We see the new quantity $\Lambda$ that introduces equilibrium anisotropy shear and curvature to the radial derivative of the normal component. We have also defined a notation for two pressure terms, `fast pressure' $p_f$ associated with cross-field compression, and `slow pressure' $p_s$ associated with parallel compression. 
\subsection{Homogenous dispersion relation}
It is useful here to set all derivatives to zero and obtain the homogenous dispersion relation. Then, the requirement for a solution of \Equation{cylinderWave} is that the determinant of the coefficient matrix vanishes.  The three branches of solutions for parallel and perpendicular propagation are
\begin{eqnarray}
\frac{\omega}{\kpar}=\pm \frac{B}{\sqrt{\rho_0}}  && \text{Shear~Alfv}\acute{\text{e}}\text{n~mode}\\
\frac{\omega}{\kpar}=\pm \sqrt{\frac{p_s}{\rho_0}}=\pm \sqrt{\frac{\frac{2}{3}\pperp +\ppar}{\rho_0}} &&\text{Isotropic~slow~ion~sound~mode } \\
\frac{\omega}{\kperp}=\pm \sqrt{\frac{p_f}{\rho_0}+\frac{B^2}{\rho_0}} \nonumber \\
=\pm\sqrt{\frac{\frac{4}{3}\pperp+\frac{1}{3}\ppar}{\rho_0}+\frac{B^2}{\rho_0}} &&\text{Isotropic~fast~magnetosonic~mode}
\end{eqnarray}
which correspond exactly to the MHD wave branches in the isotropic limit. The shear \Alfven ~wave dispersion relation is completely unchanged with all anisotropy changes contained in the magnetic geometry of the equilibrium. Looking at the remaining branches, the labelling of $p_f$ and $p_s$ has a clear interpretation in terms of the equilibrium pressure contribution to the fast and slow compressional modes.  In the isotropic limit, the pressure terms are degenerate and correspond to $\gamma p$. Conversely, an anisotropic equilibrium will see a modification in the sound speed which is a simple and measurable consequence of the SA model. One may go further and compare the single-adiabatic model alongside the double-adiabatic model, whose dispersion relations are ~\cite{Volkov1966} \cite{Thompson1961}
\begin{eqnarray}
\frac{\omega}{\kpar}=\pm  \frac{B}{\sqrt{\rho_0}} \sqrt{\Brackets{1-\Delta}}  && \text{Shear~Alfv}\acute{\text{e}}\text{n~mode}\\
\frac{\omega}{\kpar}= \pm \sqrt{\frac{3\ppar}{\rho_0}} &&\text{Slow~ion~sound~mode} \\
\frac{\omega}{\kperp}=\pm\sqrt{\frac{2\pperp}{\rho_0}+\frac{B^2}{\rho_0}} &&\text{Fast~magnetosonic~mode}.
\end{eqnarray}
The CGL model may be considered exact for an adiabatic homogenous collisionless plasma, since there is no mechanism by which the orthogonal degrees of freedom are coupled, and indeed, CGL agrees with the guiding centre plasma model under these circumstances \cite{Choe1977}. 
This comparison lays bare exactly what is wrong with the MHD and SA models through the neglect of anisotropic forces in the momentum equation for collisionless plasma.  The modification to the \Alfven~speed produced by large anisotropy and resulting firehose instability is lost. The perpendicular pressure is involved in doing parallel work and vice-versa. For non-slab geometry, these degrees of freedom should indeed be coupled, but the quantitative degree of coupling will be geometry dependent, and not strictly isotropic as we have assumed in MHD and the single-adiabatic model. The only satisfactory elimination of this assumption is by restoring the anisotropic perturbation $\widetilde{\pi}$ with non-local closures which take into account the toroidal geometry, bringing the fluid results closer to the kinetic approach.
\subsection{Cylindrical continuous spectrum}
In this final section, we write the full ordinary differential equation in the radial displacement, to allow easy exploitation and comparison with simple experiments, to compare alongside the MHD version, and to obtain the single-adiabatic cylindrical continuum. Again, adopting the notation of \cite{Goedbloed2004} for side-by-side comparison, the eigenvalue problem is
\begin{eqnarray}
\fl \Fullby r \Brackets{\frac{N\chi'(r)}{rD}}+\Fullby r\Brackets{\frac{\Lambda \rho_0 \omega^2\Brackets{\rho_0\omega^2-B^2\kpar^2}}{rD}\chi(r)}+ \Biggl[\frac{1}{r}\Brackets{\rho_0\omega^2-B^2\kpar^2} \nonumber\\
\fl- \Brackets{\frac{B_\theta^2}{r^2}}'-\frac{4k^2B_\theta^2 B^2(\rho_0\omega^2-\kpar^2p_s)+\Lambda \rho_0\omega^2 r 2kB_\theta B\kperp}{r^3D} \nonumber \\
\fl +
\Biggl( \frac{2kB_\theta B \kperp\Brackets{(p_f +B^2)\rho_0\omega^2-p_sB^2\kpar^2}}{r^2D}\Biggr)'\Biggr]\chi(r)=0 \\
\chi\equiv r \xi_n \\
N\equiv\Brackets{\rho_0\omega^2-B^2\kpar^2}\Brackets{\kpar^2p_sB^2-\rho_0\omega^2B^2-p_f\rho_0\omega^2} \\
D\equiv \Brackets{\kpar^2+\kperp^2}\Brackets{\kpar^2-\rho_0\omega^2}B^2+\rho_0\omega^2\Brackets{\rho_0\omega^2-p_s\kpar^2-p_f\kperp^2}
\end{eqnarray}
with new equilibrium curvature terms in $\Lambda$ and the `real' and `apparent' singularities of MHD spectral theory easily identified as $N=0$ and $D=0$ respectively.
The $N=0$ singular solutions express the continuous spectrum
\begin{eqnarray}
\frac{\omega}{\kpar} = \pm \frac{B}{\sqrt{\rho_0}} && \text{Shear~Alfv}\acute{\text{e}}\text{n}\text{~continuum}\\
\frac{\omega}{\kpar}=\pm \sqrt{\frac{p_s B^2}{\rho_0\Brackets{p_f+B^2}}} \nonumber \\
 =\pm \sqrt{\frac{\Brackets{\frac{2}{3}\pperp +\ppar}B^2}{\rho_0\SBrackets{\Brackets{\frac{4}{3}\pperp+\frac{1}{3}\ppar}+B^2}}}&&\text{Ion~sound~continuum}
\end{eqnarray}
directly corresponding to the MHD continuum branches, but with the breaking of pressure degeneracy into $p_f$ and $p_s$. Depending on the value of $\beta$, these changes in continuua should also be measurable.
\section{Conclusion}
In this paper, we use a fluid approach to partially address the problem of non-thermal anisotropic fast particle pressure on linear stability. This approach utilises a novel single-adiabatic closure designed to exactly reproduce the MHD stability analysis in the isotropic limit which cannot be said of the alternative CGL model, due to the anisotropic nature of collisionless modes.  The single adiabatic model considers only the `average pressure' and isotropic forces which underlie any possible fluid model in the first order drift limit. Thermodynamically, we have relaxed the condition of complete insulation of degrees of freedom that exists in CGL and seek to characterise only the isotropic part of the motion. The anisotropic fast-particle terms normally associated with stabilisation do not effect the SA terms in the model and can be considered as a separate problem requiring its own non-local closure.

We also describe the modified linear stability relations which are to be included in a MISHKA-like code. We anticipate that both CGL and the single-adiabatic closure will be implemented in that code for comparison with experiment, and for later elaboration with kinetic or two-fluid additions where the phenomena requires. A non-local closure is required to include the anisotropic corrections to the perturbed forces and this is left for future work.

A distinct advantage of a fluid model is that anyone may easily calculate the implications of the model on quantities such as the sound speed and compare directly with experiment. Although the kinetic theory is the foundation on which all other plasma models are judged, the clarity and intuition gained by fluid models in the historical development of plasma physics should not be readily discarded.
\section{Acknowledgements}
The first author would very much like to thank Sarah Newton (CCFE), Per Helander (IPP) and Jim Hastie (CCFE) for useful discussions.
This work was funded by the Australian Research Council through Grant Nos. DP1093797 and FT0991899.
This project has received funding from the European Union's Horizon 2020 research and innovation programme under grant agreement number 633053 and from the RCUK Energy Programme [grant number EP/I501045]. To obtain further information on the data and models underlying this paper please contact PublicationsManager@ccfe.ac.uk. The views and opinions expressed herein do not necessarily reflect those of the European Commission.
\section*{References}
\bibliographystyle{unsrt}
\bibliography{MIKpaper}

\begin{thebibliography}{10}

\bibitem{Lao1985}
L~L Lao, H~St. John, R~D Stambaugh, A~G Kellman, and W~Pfeiffer.
\newblock {Reconstruction of current profile parameters and plasma shapes in
  tokamaks}.
\newblock {\em Nuclear Fusion}, 25(11):1611--1622, 1985.

\bibitem{HUYSMANS1991}
G.T.A Huysmans, J.~P. Goedbloed, and W.~Kerner.
\newblock {Isoparametric Bicubic Hermite Elements for Solution of the
  Grad-Shafranov Equation}.
\newblock {\em International Journal of Modern Physics C}, 02(01):371--376,
  March 1991.

\bibitem{Hirshman1991}
S.P Hirshman and O~Betancourt.
\newblock {Preconditioned descent algorithm for rapid calculations of
  magnetohydrodynamic equilibria}.
\newblock {\em Journal of Computational Physics}, 96(1):99--109, September
  1991.

\bibitem{Fitzgerald2013}
Michael Fitzgerald, L.C. Appel, and M.J. Hole.
\newblock {EFIT tokamak equilibria with toroidal flow and anisotropic pressure
  using the two-temperature guiding-centre plasma}.
\newblock {\em Nuclear Fusion}, 53(11):113040, November 2013.

\bibitem{Qu2014}
Z~S Qu, M~Fitzgerald, and M~J Hole.
\newblock {Analysing the impact of anisotropy pressure on tokamak equilibria}.
\newblock {\em Plasma Physics and Controlled Fusion}, 56(7):075007, July 2014.

\bibitem{Cooper2009}
W~A Cooper, S~P Hirshman, P~Merkel, J~P Graves, J~Kisslinger, H~F~G Wobig,
  Y~Narushima, S~Okamura, and K~Y Watanabe.
\newblock {Three-dimensional anisotropic pressure free boundary equilibria}.
\newblock {\em Computer Physics Communications}, 180(9):1524--1533, September
  2009.

\bibitem{Lauber2013}
Philipp Lauber.
\newblock {Super-thermal particles in hot plasmas—Kinetic models, numerical
  solution strategies, and comparison to tokamak experiments}.
\newblock {\em Physics Reports}, 533(2):33--68, December 2013.

\bibitem{Ramos2005}
J.~J. Ramos.
\newblock {Fluid formalism for collisionless magnetized plasmas}.
\newblock {\em Physics of Plasmas}, 12(5):052102, 2005.

\bibitem{Chew1956}
G.~F. Chew, M.~L. Goldberger, and F.~E. Low.
\newblock {The Boltzmann Equation and the One-Fluid Hydromagnetic Equations in
  the Absence of Particle Collisions}.
\newblock {\em Proceedings of the Royal Society A: Mathematical, Physical and
  Engineering Sciences}, 236(1204):112--118, July 1956.

\bibitem{Bernstein1958}
I~B Bernstein, E~A Frieman, M~D Kruskal, and R~M Kulsrud.
\newblock {An energy principle for hydromagnetic stability problems}.
\newblock {\em Proceedings of the Royal Society of London. Series A.
  Mathematical and Physical Sciences}, 244(1236):17, 1958.

\bibitem{Kruskal1958}
M.~D. Kruskal and C.~R. Oberman.
\newblock {On the Stability of Plasma in Static Equilibrium}.
\newblock {\em Physics of Fluids}, 1(4):275, 1958.

\bibitem{Rosenbluth1959}
M.~N. Rosenbluth and N.~Rostoker.
\newblock {Theoretical Structure of Plasma Equations}.
\newblock {\em Physics of Fluids}, 2(1):23, 1959.

\bibitem{Spies1974}
GO~Spies and DB~Nelson.
\newblock {Sufficient stability criteria for plasma equilbria with tensor
  pressure}.
\newblock {\em Physics of Fluids}, 17:1865, 1974.

\bibitem{Connor1976}
JW~Connor and RJ~Hastie.
\newblock {Effect of anisotropic pressure on the localized magnetohydrodynamic
  interchange modes in an axisymmetric torus}.
\newblock {\em Physics of Fluids}, 1976.

\bibitem{Choe1977}
J~Choe, JA~Tataronis, and W~Grossmann.
\newblock {A comparison of MHD and guiding center plasma models}.
\newblock {\em Plasma Physics}, 117, 1977.

\bibitem{Rosenbluth1983}
MN~Rosenbluth, ST~Tsai, JW~{Van Dam}, and MG~Engquist.
\newblock {Energetic particle stabilization of ballooning modes in tokamaks}.
\newblock {\em Physical Review Letters}, 51(21):1967--1970, 1983.

\bibitem{Cooper1981}
WA~Cooper, G.~Bateman, DB~Nelson, and T.~Kammash.
\newblock {Neutral beam effects on Tokamak ballooning mode stability}.
\newblock {\em Plasma Physics}, 23:105, 1981.

\bibitem{Wang1990}
X.H. Wang and A.~Bhattacharjee.
\newblock {Ballooning stability of anisotropic, rotating plasmas}.
\newblock {\em Physics of Fluids B: Plasma Physics}, page 2346, 1990.

\bibitem{Cheng1994}
CZ~Cheng and Q~Qian.
\newblock {Theory of ballooning-mirror instabilities for anisotropic pressure
  plasmas in the magnetosphere potential}.
\newblock {\em Journal of geophysical research}, 99(A6):11193--11, 1994.

\bibitem{Bishop1985}
C.M. Bishop and R.J. Hastie.
\newblock {Stability of anisotropic-pressure tokamak equilibria to ideal
  ballooning modes}.
\newblock {\em Nuclear Fusion}, 25(10):1443--1449, October 1985.

\bibitem{Graves2003}
J.~P. Graves, O.~Sauter, and N.~N. Gorelenkov.
\newblock {The internal kink mode in an anisotropic flowing plasma with
  application to modeling neutral beam injected sawtoothing discharges}.
\newblock {\em Physics of Plasmas}, 10(4):1034, 2003.

\bibitem{Graves2005}
Jonathan~P. Graves.
\newblock {Internal kink mode stabilization and the properties of auxiliary
  heated ions}.
\newblock {\em Physics of Plasmas}, 12(9):090908, 2005.

\bibitem{Volkov1966}
TF~Volkov.
\newblock {Hydrodynamic description of a collisionless plasma}.
\newblock {\em Rev. Plasma Phys.(USSR)(Engl. Transl.)}, 1966.

\bibitem{Freidberg1982}
Jeffrey~P. Freidberg.
\newblock {Ideal magnetohydrodynamic theory of magnetic fusion systems}.
\newblock {\em Reviews of Modern Physics}, 54(3):801--902, 1982.

\bibitem{Kulsrud1983}
Russell~M Kulsrud.
\newblock {MHD Description of Plasma}.
\newblock In A~A Galeev and R~N Sudan, editors, {\em Basic Plasma Physics},
  volume~1. ELSEVIER SCIENCE BV, Amsterdam, 1983.

\bibitem{ISI:A1997YB84300003}
A~B Mikhailovskii, G~T~A Huysmans, W~O~K Kerner, and S~E Sharapov.
\newblock {Optimization of computational MHD normal-mode analysis for
  tokamaks}.
\newblock {\em PLASMA PHYSICS REPORTS}, 23(10):844--857, October 1997.

\bibitem{Braginskii1965}
S~I Braginskii.
\newblock {Transport processes in a plasma}.
\newblock {\em Reviews of plasma physics}, 1965.

\bibitem{Marshall1957}
W~Marshall.
\newblock {The kinetic theory of an ionized gas}.
\newblock Technical report, United Kingdom Atomic Energy Authority, Harwell,
  Berkshire, 1958.

\bibitem{Kaufman1960}
Allan~N. Kaufman.
\newblock {Plasma Viscosity in a Magnetic Field}.
\newblock {\em Physics of Fluids}, 3(4):610, 1960.

\bibitem{Frieman1960}
E.~Frieman and M.~Rotenberg.
\newblock {On Hydromagnetic Stability of Stationary Equilbria}.
\newblock {\em Revs. Modern Phys.}, 32(4):898--902, 1960.

\bibitem{Dobrott1970}
D~Dobrott and J~M Greene.
\newblock {Steady Flow in the Axially Symmetric Torus Using the Guiding-Center
  Equations}.
\newblock {\em Physics of Fluids}, 13:2391, 1970.

\bibitem{Goedbloed2010}
J~P Goedbloed, R~Keppens, and S~Poedts.
\newblock {\em {Advanced magnetohydrodynamics: with applications to laboratory
  and astrophysical plasmas}}.
\newblock Cambridge Univ Pr, 2010.

\bibitem{Goedbloed2004}
JP~Goedbloed and S~Poedts.
\newblock {\em {Principles of magnetohydrodynamics: with applications to
  laboratory and astrophysical plasmas}}.
\newblock Cambridge University Press, Cambridge, 2004.

\bibitem{Thompson1961}
W~B Thompson.
\newblock {The dynamics of high temperature plasmas}.
\newblock {\em Reports on Progress in Physics}, 24(1):363--424, January 1961.

\end{thebibliography}
\end{document}